\documentclass[12pt,a4paper]{article}
\newcommand{\bmath}[1]{\mbox{\boldmath ${#1}$}}

\begin{document}
\baselineskip 4ex
\vspace*{-2.5cm}
\hfill{\bf TSL/ISV-2000-0233}
\vspace{0.5cm}

\begin{center}
\vspace*{10mm}
{\Large{\bf Production of isoscalar pion pairs in the}\\[1ex]
{\Large $\bmath{p\,d\:\to\:^3}${\bf He}$\bmath{\,\pi\,\pi}$}
{\bf reaction near threshold}}
\\[7ex]
{\large G{\"{o}}ran F\"{a}ldt}$^{\,a,\:}$\footnote{Electronic address: 
faldt@tsl.uu.se},
{\large Anders G{\aa}rdestig}$^{\,a,\:}$\footnote{Electronic address: 
grdstg@uclapp.physics.ucla.edu}$^,\,$\footnote
{Current address: Department of Physics \& Astronomy, UCLA, Box 951547,\\
\phantom{xxx}Los Angeles, CA 90095-1547, USA},
{\large Colin Wilkin}$^{\,b,\:}$\footnote{Electronic address: cw@hep.ucl.ac.uk}
\\[1ex]
$^a$ Division of Nuclear Physics,  Box 535, 751 21 Uppsala, Sweden
\\[1ex]
$^b$  Physics and Astronomy Department, UCL, London, WC1E 6BT, UK\\[3ex]
\today\\[4ex]
\end{center}

\begin{abstract}
The production near threshold of isoscalar pion pairs in the 
$p\,d\:\to\:^3$He$\,(\pi\,\pi)^0$ reaction is estimated in a two-step model
which successfully describes the production of $\eta$, $\omega$ and $\eta'$ 
mesons. A virtual pion beam, generated through an $N\,N\to d\,\pi$ reaction
on one of the nucleons in the deuteron, produces a second pion \textit{via}
a $\pi\, N\to \pi\,\pi\, N$ reaction on the other nucleon. Using the same
scale factor as for heavy meson production, the model reproduces the total 
$\pi^0\,\pi^0$ production rate determined at an excess energy of 37~MeV.
There are some indications in the data for a suppression of events with
low $\pi\pi$ masses, as in the $\pi^-\,p\to \pi^0\,\pi^0\,n$ reaction, and
this is confirmed within the model.
The model suggests that a significant fraction of the charged pion
production in the $p\,d\:\to\:^3$He$\,\pi^+\,\pi^-$ reaction at $Q=70$~MeV
might be associated with isoscalar pion pairs, though this does not explain
the strong dependence observed on the $\pi^+\,\pi^-$ relative momentum angle.
\end{abstract}

\vspace{5mm}


\vfill
\baselineskip 2ex
\noindent
{Corresponding author:\\
Colin Wilkin,\\
Physics \& Astronomy Dept.,
UCL, Gower St.,
London WC1E 6BT.}

\newpage

\baselineskip 5ex

The study of neutral two-pion production through the $p\,d\:\to\:^3$He$\,X^0$ 
reaction has a long history. At excess energies $Q$ (the c.m.\ kinetic energy 
in the final state) around 200-300~MeV, sharp structure is seen at 
missing masses of about 310~MeV/c$^2$~\cite{ABC,Banaigs}. The absence of any
significant strength in the $p\,d\:\to\:^3$H$\,X^+$ channel at low $m_X$ means 
that the effect is associated with isospin-zero pion-pion pairs which,
because of the available energy, must be dominantly in $s$-waves.
Although a quantitative explanation of the ABC enhancement has not yet
been provided for this reaction,
a similar effect in $n\,p\to d\,X^0$ has been shown to originate from
the excitation of two $\Delta$-isobars~\cite{Risser}. The prominent ABC
peaks in the $d\,d\:\to\:^4$He$\,X^0$ case have also been shown to be due
to double pion $p$-wave production~\cite{GFW}. 

The experimental picture changes dramatically at lower energies. For values
of $Q$ around 70-90~MeV, the angular distributions observed by the
MOMO group~\cite{MOMO1,MOMO2} for the exclusive 
$p\,d\:\to\:^3$He$\,\pi^+\,\pi^-$ reaction suggest strongly that the
$\pi^+\,\pi^-$ spectrum is mainly $p$-wave in nature, and hence has
isospin-one. The measurement of the $\pi^0\pi^0/\pi^+\pi^-$ charge ratio at
CELSIUS at an excess energy (with respect to the $\pi^0\pi^0$
threshold) of $Q= 37$~MeV~\cite{Kat} shows that there is significant $I=1$
production even at this much smaller $Q$. The
isoscalar $\pi^0\,\pi^0$ spectrum, determined either by direct 
measurement~\cite{Kat} or through the subtraction of 
exclusive $^3$He$\,\pi^+\,\pi^-$
data from an inclusive $p\,d\:\to\:^3$He$\,X^0$ measurement~\cite{MOMO2},
shows that there is no $s$-wave ABC enhancement at low $Q$. On the
contrary, there are rather indications that the $s$-wave cross section
is actually suppressed at low $\pi^0\,\pi^0$ masses as compared to phase
space~\cite{Kat}. It is the aim of the present paper to
demonstrate that near-threshold isoscalar two-pion production in the
$p\,d\:\to\:^3$He$\,\pi\,\pi$ can be described in terms of
sequential single-pion production.

\noindent
\input epsf
\begin{figure}[h]
\begin{center}
\mbox{\epsfxsize=4in \epsfbox{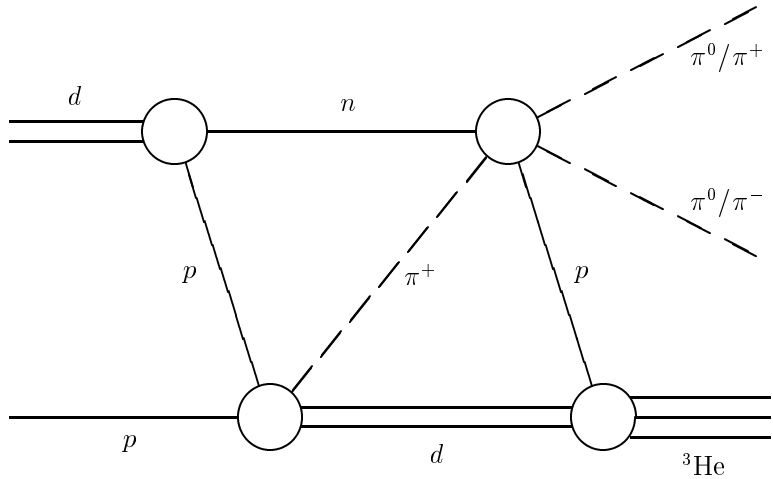}}
\caption{Dynamical model for the $p\,d\:\to\:^3$He$\,(\pi\,\pi)^0$ reaction
in terms of sequential $pp\to d\,\pi^+$ and $\pi^+\,n\to (\pi\,\pi)^0\, p$
processes. There is an analogous contribution from intermediate
neutral pions.} 

\end{center}
\label{fig1}
\end{figure}

The large momentum transfers required to produce heavy mesons, such as the
$\eta$ or $\omega$, through the $p\,d\:\to\:^3$He$\,X^0$ reaction mean
that two-step processes which minimise the momentum mismatch in the
nuclear wave functions can provide the dominant driving force. In one
such model, a pion produced on one of the nucleons in the target deuteron
is converted into the observed heavy meson through an interaction on the
second of the target nucleons~\cite{FW1}. Apart from an {\it ad hoc} overall
normalisation factor $N\approx 2.4$, which may reflect the retention
of only {\it bound} intermediate deuteron states in the calculation, this
approach describes
well the threshold amplitudes for producing $\eta$, $\omega$ and $\eta'$,
though the experimental $\phi$ yield is a
little too high~\cite{FW2}. This success may be attributed to the fact that
the intermediate pion in the diagram is close to its mass shell.
We wish to apply the same model to isoscalar $\pi\,\pi$
production by introducing rather a final $\pi\,N\to (\pi\,\pi)^0\,N$ process,
as in Fig.~1. For definiteness, we consider $\pi^0\,\pi^0$ production;
estimates of charged pion production in the $I=0$ channel will then follow
from isospin invariance, after correcting for the pion mass difference.

The principal difference with the earlier work~\cite{FW2} is that the
low mass $\pi^0\,\pi^0$ system is in a $0^+$ state and so there is a parity
change at the $\pi^+\,n\to \pi^0\,\pi^0\,p$ vertex. Parameterising this
amplitude in terms of two-component Pauli spinors $u_{p(n)}$ as 
\begin{equation}
\label{e0}
M(\pi^+n\to \pi^0\pi^0p)= a(m_{\pi\pi},W_{\pi N})\,u_p^{\dagger}\,
\bmath{\sigma}\cdot\bmath{p}_{\pi}\,u_{n}\:,
\end{equation}
the corresponding differential cross section for $s$-wave production is
\begin{equation}
\label{e1}
\textrm{d}\sigma(\pi^+n\to \pi^0\pi^0p) = 
\textrm{d}\sigma(\pi^-p\to \pi^0\pi^0n)
= \frac{1}{64\pi^3}\,\frac{p\,p'}{W_{\pi N}^2}\,|a(m_{\pi\pi},W_{\pi N})|^2\,
k_{\pi}^*\,\textrm{d}m_{\pi\pi}\:.
\end{equation}
Here $p$ and $p'$ are the incident and final nucleon momenta in the 
overall c.m.\ system where the total energy is $W_{\pi N}$. In the
$\pi\pi$ rest frame, $k_{\pi}^*$ is the relative momentum, which is
related to the $\pi\pi$ invariant mass through 
$m_{\pi\pi}=2\sqrt{k_{\pi}^{*\,2}+m_{\pi}^2}$. We shall neglect the angular
dependence of the amplitudes in the present work.

Because of the nature of the two-step process in Fig.~1, only small Fermi 
momenta are required. Working to first order in these momenta, as in 
\cite{FW1}, we find that the amplitudes are proportional to the complex form
factors
\begin{equation}
\label{e2}
S_{\alpha\beta}(\mathbf{W},\mathbf{V})= 
(2\pi)^3 \int_{0}^{\infty} \textrm{d}t\,e^{it\Delta E_{0}} 
\psi_{\alpha}^{*}(-t\mathbf{W})\,\varphi_{\beta}(t\mathbf{V})\:.
\end{equation}
These involve integrals over configuration--space deuteron $(\varphi_{\beta})$
and $^3$He $(\psi_{\alpha})$ wave functions, where $\alpha,\beta=(0,2)$
represent nuclear $S$-- and $D$--state components.
The energy mismatch $\Delta E_{0}$ between the intermediate and external
energies for zero Fermi momenta is generally small for near-threshold 
heavy meson production in this model.

The relativistic relative velocity vectors $\mathbf{V}$ and $\mathbf{W}$, 
\begin{eqnarray}
\label{e3}
\mathbf{V} 
&=&\frac{2}3\frac{1}{E_{\pi}(\frac{2}3\mathbf{p}_{\pi\pi}-
\frac{1}{2}\mathbf{p}_d)} 
\,\mathbf{p}_{\pi\pi}-\frac{1}{2}\left[ \frac{1}{E_{\pi}
(\frac{2}3\mathbf{p}_{\pi\pi}-
\frac{1}{2}\mathbf{p}_d)}+
\frac{1}{E_{n}(\frac{1}{2}\mathbf{p}_d)}\right]\mathbf{p}_d\:,
\\[1ex] 
\nonumber
\mathbf{W}
&=&-\frac{2}{3}\left[ \frac{1}{E_{\pi}(\frac{2}3\mathbf{p}_{\pi\pi}-
\frac{1}{2}\mathbf{p}_d)}+\frac{1}{E_d(-\frac{2}3\mathbf{p}_{\pi\pi})}\right]
\mathbf{p}_{\pi\pi}+\frac{1}{2} \frac{1}{E_{\pi}(\frac{2}3\mathbf{p}_{\pi\pi}-
\frac{1}{2}\mathbf{p}_d)} \,\mathbf{p}_d\:,
\end{eqnarray}
where $\mathbf{p}_{\pi\pi}$ is the total $\pi\pi$ momentum vector and
$\mathbf{p}_{d}$ that of the initial deuteron in the overall c.m.\ frame.
The component of $\mathbf{V}$ along $\mathbf{p}_d$ must be subjected to a
Lorentz contraction~\cite{FW1}. The relativistic energies $E_i$ are 
evaluated at the values of the momenta indicated.

For zero Fermi momenta, the $pp\to d\,\pi^+$ amplitudes should be evaluated
in the forward direction for threshold heavy meson production. The forward 
direction assumption is also very good even away from threshold provided that
$p_{d} \gg p_{\pi\pi}$, as it is in cases under investigation. There 
are two $pp\to d\,\pi^+$ amplitudes in the forward direction but, at the
energies required here, the helicity-zero completely dominates over the 
helicity-one~\cite{SAID}. Keeping then only the dominant amplitude $A$, the
c.m.\ differential cross section is 
\begin{equation}
\label{e4}
\frac{\textrm{d}\sigma}{\textrm{d}\Omega}(pp\to d\,\pi^+)
= \frac{1}{128\pi^2}\,\frac{p_{\pi}}{p_pW_{pp}^2}\,|A|^2\:.
\end{equation}

The evaluation of the unpolarised differential cross section in this model is 
similar to that for the production of single heavy mesons~\cite{FW1}
and leads to
\begin{eqnarray}
\nonumber
\textrm{d}\sigma(pd\to \: ^3\textrm{He}\,\pi^0\pi^0)
= \frac{p_{\pi\pi}}{p\,W_{pd}^2\,m_p^2\,E_{\pi}(\frac{2}3\mathbf{p}_{\pi\pi}-
\frac{1}{2}\mathbf{p}_d)^2}\,\frac{9}{2^{21}\pi^{10}}\,N\,
\left\{|S_a|^2+|S_b|^2\right\}\,\\[1ex]
\label{e5}
\times\:|A|^2\,|a(m_{\pi\pi},W_{\pi N})|^2\,
k_{\pi}^*\,p_{\pi}^2\,\textrm{d}m_{\pi\pi}\:
\textrm{d}\Omega_{\textrm{\scriptsize He}}\:,
\end{eqnarray}
where an isospin factor of $\frac{9}{4}$ has been included to account 
for the $\pi^0$-exchange term in Fig.~1.
The form factor combinations required are
\begin{equation}
\label{e6}
S_a= S_{00}-S_{20}\sqrt{2}\:,\hspace{1cm}
S_b= S_{02}-S_{22}\sqrt{2}\:.
\end{equation}
In order to describe
$(\eta,\,\omega,\,\eta',\,\phi)$ production, it was found necessary to
multiply the analogous prediction by a normalisation factor
$N=2.4$~\cite{FW1}.

Single pion production in pion-nucleon collisions has been measured in 
many charge states near threshold and parameterisations given for the
total cross sections as functions of the beam energy~\cite{Hugh}.
The charge dependence indicates that the cross section is dominated by
$I=0$ pion pairs for $Q< 100$~MeV. This is consistent with the smallness of
the anisotropy in the angular distribution of the $\pi^+\,\pi^-$ relative
momentum for $\pi^-\,p\to \pi^+\,\pi^-\,n$, which arises from $s$-$p$,
and hence $I=0/I=1$ interference~\cite{Kermani}. Data on
the $\pi^-\,p\to \pi^0\,\pi^0 n$ reaction in the $Q\approx 50-100$~MeV 
region show clear evidence for the suppression of events with low 
$m_{\pi\pi}$~\cite{Lowe}. This is also seen for $\pi^-\,p\to \pi^+\,\pi^-\,n$
but not $\pi^+\,p\to \pi^+\,\pi^+\,n$, where only $I=2$ pion pairs are
produced~\cite{Kermani}.
In the dynamical model of the Valencia
group~\cite{Oset}, this shift towards higher $m_{\pi\pi}$ is due to an
accidental cancellation between two contributions, one of which involves
the double pion $p$-wave decay of the Roper resonance 
$N^*(1440)\to N\,\pi^0\,\pi^0$.

Although the production of isospin-two $\pi\pi$ pairs is small~\cite{Kermani},
it has to be subtracted from the $\pi^0\pi^0$ data to get the $I=0$ rate
required in Eq.~(\ref{e5}) This subtraction is model-dependent and, for
this purpose, we have used the predictions of the Valencia model~\cite{Oset},
which describes reasonably well the shape of the experimental 
data~\cite{Lowe,Kermani}. A global fit to their $I=0$ predictions,
renormalised slightly to agree with the overall $\pi^0\pi^0$ amplitude
analysis of Lowe and Burkhardt~\cite{Hugh}, gives
\begin{equation}
 \frac{1}{64\pi^3}\,|a(m_{\pi\pi},Q)|^2 =
(1.092-0.0211Q+0.00015Q^2)
\label{e7}
\end{equation}
\[+(4.18+0.0075Q-0.00098Q^2)x
+(47.65-0.935Q+0.00743Q^2)x^2\:\mu\textrm{b/MeV}^2\:, 
\]
which is valid up to
$Q'=Q\,(1+m_{\pi}/m_{\textrm{\scriptsize He}})^2/(1+m_{\pi}/m_{p})^2
\approx 100$~MeV. Given that $x=m_{\pi\pi}/m_{\pi}-2$,
this illustrates the suppression of the
matrix element at low $m_{\pi\pi}$, a feature which becomes even
more pronounced at higher $Q$. 

\input epsf
\begin{figure}[ht]
\begin{center}
\mbox{\epsfxsize=4in \epsfbox{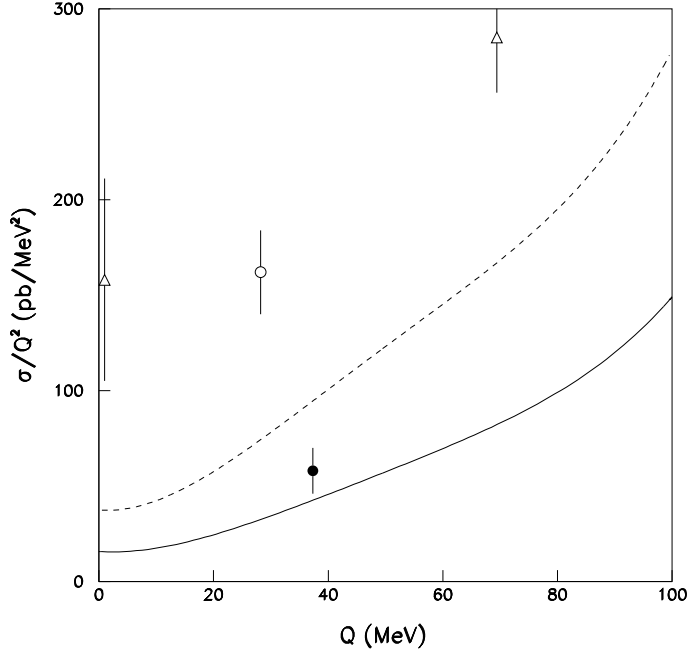}}
\caption{Total cross sections for the $p\,d\:\to\:^3$He$\,(\pi\,\pi)^0$
reactions,
divided by $Q^2$ as functions of the excess energy $Q$. The predicted 
solid and broken curves refer to $I=0$ $\pi^0\,\pi^0$ and $\pi^+\,\pi^-$
production respectively, as do the closed and open circles from 
CELSIUS~\protect\cite{Kat}. The triangle is the published MOMO $\pi^+\,\pi^-$
data point~\protect\cite{MOMO1}. The near-threshold IUCF~\protect\cite{Vigdor}
$\pi^+\,\pi^-$ point (square) is strongly influenced by Coulomb distortion.}
\end{center}
\label{fig2}
\end{figure}

Our predictions for the $p\,d\:\to\:^3$He$\,(\pi\,\pi)^0$
total cross section divided by $Q^2$, obtained using the same value of
$N=2.4$ which gave good agreement for heavy meson production, are to
be found in Fig.~2.
The steady increase with $Q$ is mainly a reflection of the energy
dependence of the $pp \to d\,\pi^+$ and $\pi^-\,p\to \pi\,\pi\,n$ amplitudes;
the average form factor changes comparatively little. The solid curve
passes close to the CELSISUS $\pi^0\,\pi^0$ point~\cite{Kat}, but the broken
one is significantly too low, indicating the presence of some $I=1$ 
$\pi^+\,\pi^-$ production. The IUCF point
was obtained at $Q=0.67$~MeV~\cite{Vigdor}, and hence can be assumed to be
purely $s$-wave, though it is heavily influenced by Coulomb effects. The
comparison of our total cross section predictions with the MOMO $\pi^+\,\pi^-$
point at $Q=70$~MeV~\cite{MOMO1} would suggest that it is mainly $I=0$ pairs
which are being produced, though this is at variance with the strong
dependence observed on the angle of the $\pi^+\,\pi^-$ relative momentum.

The $I=0$ $m_{\pi\pi}$ distributions expected at the CELSIUS energy are
illustrated in Fig.~3 and these demonstrate the shift to higher masses as
compared to phase space, which is apparent in the $\pi\,N\to\pi\,\pi\,N$
input. These experimental data~\cite{Kat} have insufficient statistics to
draw definitive conclusions on the shape of the spectrum. It should be noted
that the CELSIUS integrated cross section points shown in Fig.~2 are mainly
determined by the higher statistics of their inclusive measurement,
which was carried out simultaneously~\cite{Kat}. On the other hand,
a low $m_{\pi\pi}$ suppression in the 70~MeV MOMO data is clear
in their high statistics exclusive $\pi^+\pi^-$ production results
shown in Fig.~4.
\input epsf
\begin{figure}[ht]
\begin{center}
\mbox{\epsfxsize=4in \epsfbox{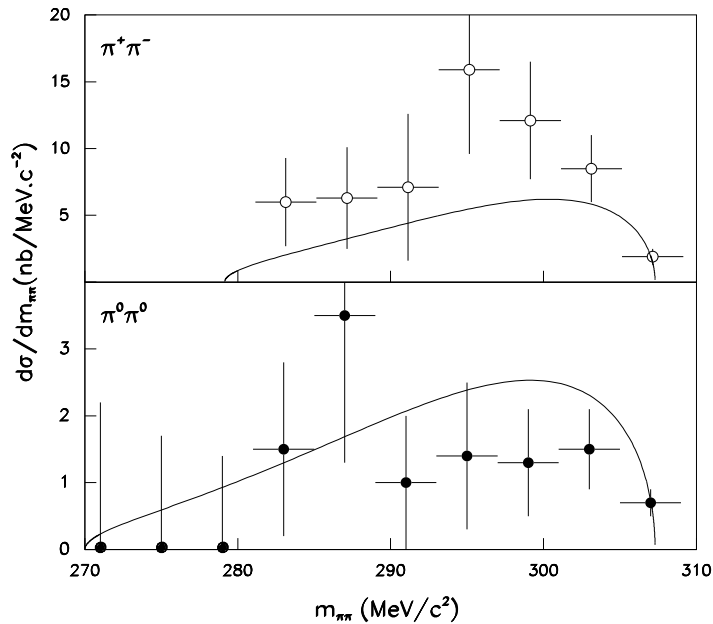}}
\caption{Predicted $I=0$ $\pi\pi$ effective mass distributions for the 
$p\,d\:\to\:^3$He$\,\pi^+\,\pi^-/\pi^0\,\pi^0$ reactions at an
incident energy of 477~MeV compared with the CELSIUS experimental
data~\protect\cite{Kat}.}
\end{center}
\label{fig3}
\end{figure}

\input epsf
\begin{figure}[ht]
\begin{center}
\mbox{\epsfxsize=4in \epsfbox{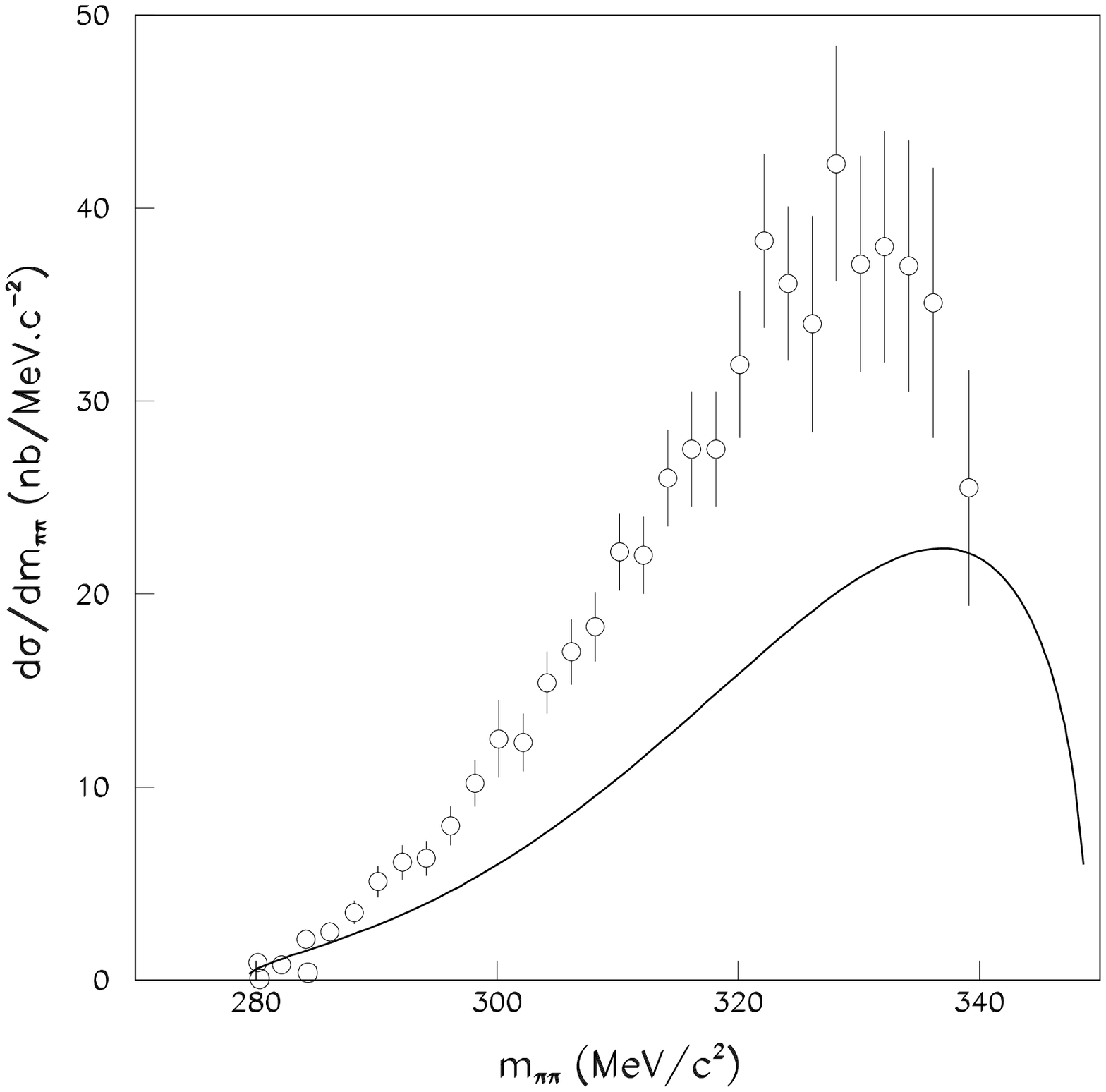}}
\caption{Predicted $I=0$ $\pi^+\,\pi^-$ effective mass distributions for the 
$p\,d\:\to\:^3$He$\,\pi^+\,\pi^-$ reactions at an
incident energy of 546~MeV, compared with the COSY experimental
data~\protect\cite{MOMO1}.}
\end{center}
\label{fig4}
\end{figure}

The MOMO data~\cite{MOMO1} show a strong dependence upon the angle
$\theta_{\pi\pi p}$ between the relative $\pi\,\pi$ momentum and that of
the beam direction. Taken together with the suppression of events at low
$m_{\pi\pi}$, this suggests the production of $I=1$, $\ell=1$ $\pi\pi$
pairs with spin projection $m=\pm 1$ along the beam direction. Such an
interpretation is backed by the group's preliminary data on the inclusive
$p\,d\:\to\:^3$He$\,X^0$ reaction~\cite{MOMO2}, which indicate a
$\pi^0\,\pi^0$ production rate less than half of that predicted in Fig.~2.

The production of $I=0$, $\ell=0$ $\pi\pi$ pairs would give no dependence
upon $\theta_{\pi\pi p}$, though an interference with an $I=0$,
$\ell =2$ contribution could lead to such a variation. However, there is no
sign of any effect of this kind in
$\pi^-\,p\to \pi^+\,\pi^-\,n$~\cite{Kermani}.
Since $\pi^+\,\pi^-$ $p$-waves are so small in
$\pi^-\,p\to \pi^+\,\pi^-\,n$ near threshold~\cite{Kermani}, any simple
extension of our model to include $p$-wave production cannot lead to
$\pi\pi$ $p$-wave dominance.

For two-pion production near threshold, the intermediate pion in Fig.~1
gets closer to its mass shell when the $^3$He emerges along the direction
of the initial proton beam and this increases the magnitude of the
average form factor. The two-step model therefore predicts that, for
low $m_{\pi\pi}$, the dipion should be produced preferentially in the
backward hemisphere. This effect will, of course, disappear at high masses
because the situation then approaches one of near-threshold kinematics.

We have shown that the gross features found in the production of isoscalar 
pion pairs in the $p\,d\:\to\:^3$He$\,(\pi\,\pi)^0$ reaction near
threshold can be understood in terms of the creation of an intermediate
virtual pion beam, which in turn produces a second meson. The only way that 
such a model could generate an ABC peak in the $Q\approx 250$~MeV region 
is if this were already present in the  $\pi^-\,p\to \pi^0\,\pi^0\,n$ input. 
The parameterisation of the results of the Valencia model in Eq.~(\ref{e7}) 
corresponds to a parabola in $m_{\pi\pi}$, whose minimum moves to higher 
values as $Q$ increases. This is due to the enhanced importance of the 
Roper contribution and may leave space at low masses for an ABC 
effect at higher $Q$. The question may soon be resolved, because 
data on $\pi^-\,p\to \pi^0\,\pi^0\,n$ at $p_{\pi}=750$~MeV/c are currently 
being analysed by the Crystal Ball collaboration~\cite{Ben}.

Further theoretical work is needed to include a more detailed description
of the $\pi\,N\to \pi\,\pi\,N$ input, though any improvement will, inevitably,
be rather model-dependent.
\\

It is a pleasure to thank R.~Jahn for extensive discussions regarding the
data of Ref.~\cite{MOMO1,MOMO2} and M.~Andersson, P.E.~T\'egner, and
K.~Wilhelmsen Rolander for those pertaining to Ref.~\cite{Kat}.
We are greatly indebted to 
M.~Vincente Vacas for giving us numerical values of the calculations 
reported in \cite{Oset} and providing the $I=0$ projections. 
This work has been supported by
the Royal Swedish Academy of Sciences within the framework of 
the European Science Exchange Programme. One of the authors (CW) is
grateful to the The Svedberg Laboratory for its generous hospitality.
\vspace{2cm}
\end{document}